# Generalization of the Concept of Bandwidth

Alireza Mojahed[1], Lawrence A. Bergman[2], Alexander F. Vakakis[3]

[1] Department of Mechanical Engineering,
Massachusetts Institute of Technology, Cambridge, MA 02139

[2] Department of Mechanical Science and Engineering, University of Illinois, Urbana, IL 61801

[3] Department of Aerospace Engineering, University of Illinois, Urbana, IL 61801

**Abstract**

In the sciences and engineering, the concept of bandwidth is often subject to interpretation depending upon context and the requirements of a specific community. The focus of this work is to formulate this concept for a general class of passive oscillatory dynamical systems, including but not limited to mechanical, structural, acoustic, electrical, and optical. Typically, the (linearized) bandwidth of these systems is determined by the half-power (-3 dB) method, and the result is often referred to as "half-power bandwidth." The fundamental assumption underlying this definition is that the system performance degrades once its power decreases by 50%; moreover, there are restrictive conditions, rarely met, that render a system amenable to the use of this approach, such as linearity, low-dimensionality, low-loss, and stationary output. Here the concept of root mean square (RMS) bandwidth is considered, justified by the Fourier uncertainty principle, to generalize the definition of bandwidth to encompass linear/nonlinear, single/multi-mode, low/high loss and time-varying/invariant oscillating systems. By tying the bandwidth of an oscillatory dynamical system directly to its dissipative capacity, one can formulate a definition based solely on its transient energy evolution, effectively circumventing the previous restrictions. Further, applications are given that highlight the limitations of the traditional half-power bandwidth; these include a Duffing oscillator with hardening nonlinearity, and a bi-stable, geometrically nonlinear oscillator with tunable hardening or softening nonlinearity. The resulting energy-dependent bandwidth computations are compatible with the nonlinear dynamics of these systems, since at low energies they recover the (linearized) half-power bandwidth, whereas at high energies they accurately capture the nonlinear physics. Moreover, the bandwidth computation is directly tied to nonlinear harmonic generation in the transient dynamics, so that the contributions to the bandwidth of the individual harmonics and of inter-harmonic targeted energy transfers can be directly quantified and studied. The new bandwidth definition proposed in this work has broad applicability and can be regarded as a generalization of the traditional linear half-power bandwidth which is used widely in the sciences and engineering.

## 1. Introduction

The term "bandwidth" frequently arises throughout the Sciences and Engineering from the macro- to the nanoscale, and in diverse areas ranging from optics, electromagnetics, and imaging to sensing, dynamics and controls. For instance, in the realm of photonics it appears in different contexts [1-4]. Light sources have definitive optical bandwidths, which refer to the widths of the optical spectra of the corresponding outputs. Narrow-linewidth lasers typically have extremely small bandwidths – as low as 1 Hz, which are orders of magnitude less than their mean optical frequencies [5]. Conversely, ultrashort pulses with femtosecond durations possess very large bandwidths – as large as tens of terahertz [6]. In such cases, bandwidth refers to the "breadth" in frequency of the corresponding optical spectrum. A similar definition is also commonly applied to the reflection bandwidth of a mirror [7], the optical transmission bandwidth of an optical fiber [8,9], the gain bandwidth of an optical amplifier [10], and the phase-matching bandwidth of a nonlinear optical device [11].

In dynamics, vibrations and acoustics, the notion of bandwidth as described above is also related to the Q-factor (quality factor) which, for a resonant linear single degree-of-freedom (SDOF) system, is an indication of the spectral sharpness of the resonance in the frequency domain [12,13]. In fact, the Q-factor equals the (natural) logarithmic decrement between two successive local maxima of a damped waveform, which is related, in turn, to the decay-rate of the wave energy, $Q = \omega_0/\Gamma$, where $\omega_0$ is the resonance frequency of the linear resonator and $\Gamma$ the decay-rate of the wave energy [12-16].

The concept of bandwidth is of special significance to the fields of control and dynamical systems. In fact, the performance of amplifiers and filters in mechanical and electromagnetic systems is measured in the frequency domain by their related bandwidths. For instance, an important feature of a bandpass filter, i.e., a device that allows for a certain range of frequencies to pass and stops (attenuates) frequencies outside that range, is its bandwidth [14]. For this specific case, the "fractional bandwidth" (to be defined below) is used to quantify the bandwidth of the filter [17,18]. In a dynamical system such as an RLC circuit, a system of linear oscillators, a fiber optical system with linear intensity modulation, and an optical resonator, the corresponding bandwidth provides a measure of the system's response decay (dissipation) rate, or as mentioned previously, its Q-factor [1-4,13,14,19].

## 2. Concept of bandwidth

We start by providing a brief overview of different definitions of bandwidth spanning different areas and applications. Then we proceed to formulate a new definition of bandwidth which is sufficiently general to apply to broad classes of linear or nonlinear oscillatory dynamical systems.

### 2.1 Current definitions

As a general definition the bandwidth of an oscillator is an inherent (intrinsic) property of the oscillator and provides a measure of time locality of the energy, i.e., the capacity of the oscillator to dissipate energy in time. In signal theory and processing, dynamical systems, control systems, optics and electromagnetics, different bandwidth definitions are employed, some of which are briefly reviewed below.

Rayleigh bandwidth: Rayleigh bandwidth is defined as the inverse of the duration of a pulse, such as in radar applications [20]. For instance, the bandwidth of a pulse with wavelength of 100 $\mu s$ is equal to 10 Hz.

Absolute bandwidth: Absolute bandwidth is defined as the difference between the upper ($\omega_h$) and the lower ($\omega_l$) frequency limits, $\Delta\omega_a = \omega_h - \omega_l$, of the spectrum of a signal $x(t)$ in the frequency domain $\omega$. To this end, the spectrum of a signal $x(t)$ is defined as $S_x(\omega) = |X(\omega)|$, where the Fourier transform of the signal is $X(\omega) = \int_{-\infty}^{\infty} x(t)e^{-j\omega t}dt$, $j = \sqrt{-1}$, and $|\bullet|$ denotes $L_1$-norm [21]. It should be noted that for this definition $S_x(\omega)$ must be perfectly band-limited.

Fractional bandwidth: Fractional bandwidth of a signal is expressed as the ratio between the absolute bandwidth, $\Delta\omega_a$, and the center frequency, $\omega_{center} = (\omega_h + \omega_l)/2$, of the signal [17,18,22,23].

Ratio bandwidth: This is defined as the ratio between the upper ($\omega_h$) and lower ($\omega_l$) frequency limits of the spectrum of a signal $x(t)$ along the frequency axis [23].

Essential bandwidth: Essential bandwidth of a signal is defined as the portion of its spectrum, i.e., the magnitude of the Fourier transform of the signal, that contains most of the energy [24].

x dB bandwidth: In this definition, the bandwidth is defined as the frequency range of the spectral density, i.e., the square of the Fourier transform magnitude of the signal, at $x$ dB relative to its maximum. Underlying this definition is the assumption that the performance of a device (or strength of a signal) degrades once its power drops beyond $x$ dB from its maximum. In its most common use, $x = -3$, which corresponds to the half-power level of the power spectrum, i.e., $x = 10\log_{10} 1/2 \sim -3.0103$ dB. This is also referred to as the half-power bandwidth, the full width at half maximum (FWHM) bandwidth, and the half width at half maximum (HWHM) bandwidth [3,25] – cf. Figure 1.

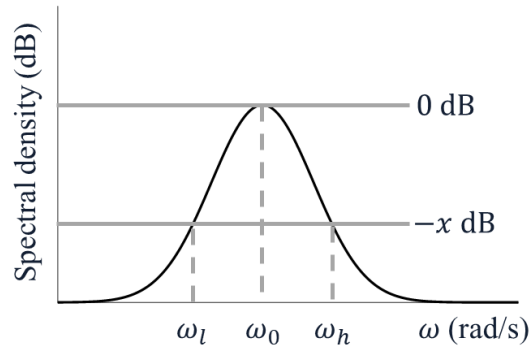

Figure 1. Definition of $x$ dB bandwidth of the spectral density, $S_{xx}(\omega) = S_x^2(\omega)$, of a generic signal $x(t)$, where $\omega_0$ is the center frequency of the signal (the definition of which is different than the center frequency defined in the case of fractional bandwidth), and $\omega_l$ and $\omega_h$ are frequencies at which the spectral density is at $-x\,dB$ relative to its maximum value; the $x\,dB$ bandwidth in this case is defined as $\Delta\omega_{x\,dB} = \omega_h - \omega_l$.

The half-power bandwidth (3 dB bandwidth) of a linear resonant device with natural frequency equal to $\omega_0$ is computed as $\Delta\omega_{-3\,dB} = \omega_0/Q$, where $Q$ is the quality factor of the resonator; the quality factor of the resonator is proportional (with a factor of $2\pi$) to the ratio of its instantaneous energy to the rate of its energy loss – the energy it loses at each period of oscillation. To this end, consider a linear single degree-of-freedom (SDOF) oscillator with governing equation,

$$\ddot{x} + 2\zeta\omega_0\dot{x} + \omega_0^2 x = f(t), \quad x(0) = 0, \dot{x}(0) = 0 \quad (1)$$

where $x$ is the displacement of the oscillator, $f(t)$ is the mass-normalized applied force, $\omega_0$ its natural frequency, $\zeta$ its critical viscous damping ratio, and overdot denotes differentiation with respect to time. Then the quality factor of the resonator is equal to $Q = 1/(2\zeta)$ and its half-power bandwidth is $\Delta\omega_{-3\,dB} = 2\zeta\omega_0$. The transfer function associated with the oscillator (1) can be expressed in the frequency domain as [14]:

$$\frac{X(\omega)}{F(\omega)} \equiv H(\omega) = \frac{1}{\omega_0^2 - \omega^2 + 2j\zeta\omega_0} \quad (2)$$

It can be shown that for weak damping, $\zeta \ll 1$, at $\omega = \omega_0$, $\max\{S_{xx}(\omega)\} = |H(\omega = \omega_0)|^2 = Q^2$ – cf. Figure 2. It must be emphasized that the aforementioned values of $Q$ and $\Delta\omega_{-3\,dB}$ are *approximate*, since they only hold in the limit of weak damping, i.e., for $\zeta \ll 1$ [12,14]. Moreover, as explained above the notion of half-power bandwidth and quality factor $Q$ are related to the damping ratio and the decay rate of the energy $(E)$ of the oscillator, where $E = \dot{x}^2/2 + \omega_0^2 x^2/2$.

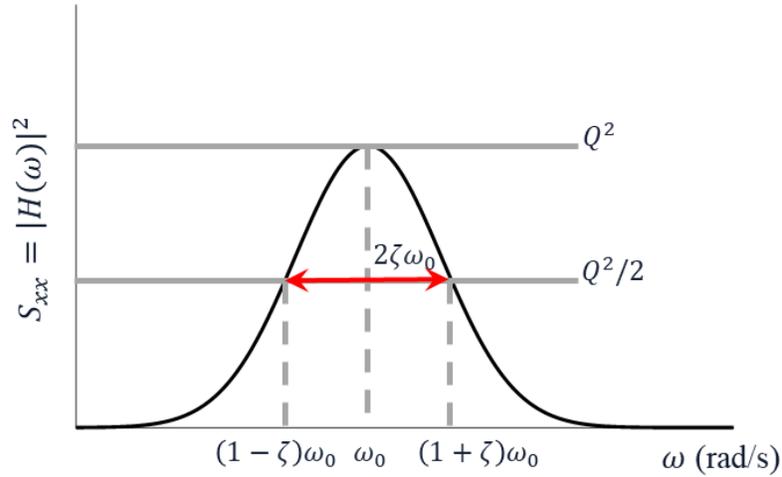

Figure 2. Spectral density of the linear SDOF oscillator (1), showing its maximum of $Q^2$, the center frequency $\omega_0$ and the half-power bandwidth equaling $2\zeta\omega_0$.

Root mean square (RMS) bandwidth: For an arbitrary but time-limited signal, $x(t)$, the Fourier transform is defined as $X(\omega) = \int_{-\infty}^{\infty} x(t) e^{-j\omega t} dt$, and the root mean square (RMS) bandwidth is defined by [26-28]:

$$\Delta\omega_{rms}^2 = \frac{4\int_0^\infty (\omega-\omega_c)^2 |X(\omega)|^2 d\omega}{\int_0^\infty |X(\omega)|^2 d\omega} \tag{3}$$

For our discussion it is important to note that relation (3) can be rewritten as,

$$\frac{\Delta\omega_{rms}}{2} = \left(\frac{m_2}{m_0}\right)^{1/2} \tag{4}$$

which is known as the Gabor bandwidth, $G$, where $m_k = \int_{-\infty}^{\infty} \omega^k |X(\omega)|^2 d\omega$ is the $k$-th moment of the spectrum (spectral moment). This specific definition of bandwidth is used when characterizing lowpass and bandpass (band-limited) filters. It should be noted that (3) is linearly proportional to the variance of $|X(\omega)|$. This is an important feature of the RMS bandwidth, to which we will refer later in the discussion.

## 2.2 Extended bandwidth definition for oscillatory dynamical systems

As noted, for typical linear time-invariant (LTI) systems, the half-power bandwidth gives an accurate representation of the decay-rate (typically the viscous damping coefficient), which holds for lightly damped SDOF oscillators. Along with this assumption, the half-power bandwidth of a SDOF linear system is based on its frequency response function (FRF) – or equivalently, the Fourier transform of its response to the unit impulse. The main drawback of this definition is that it cannot be applied to SDOF nonlinear (or multi-mode linear) oscillators, especially those with energy-dependent characteristic frequencies. This issue arises when one considers the FRF of a SDOF nonlinear oscillator, which, in contrast to its linear counterpart, depends on the amplitude of the applied force (or the energy level in general). For such signals, the Fourier transform cannot be computed since it is a linear transformation and violates the orthogonality and superposition conditions between different harmonic components [29]. We highlight this basic limitation by means of an example.

To this end, consider a lightly damped (normalized) Duffing oscillator with hardening stiffness nonlinearity, and governing equation with normalized parameters and variables expressed as,

$$u''(\tau) + \varepsilon\lambda u'(\tau) + u(\tau) + \varepsilon u^3(\tau) = 0 \tag{5}$$

where $u$ denotes normalized displacement, $\tau$ normalized time, $\lambda$ the damping coefficient, prime differentiation with respect to $\tau$, and $0 < \varepsilon \ll 1$ a small parameter. To compute the half-power bandwidth of (5), first we need to compute its FRF under the condition of primary resonance. To do this we consider an applied harmonic excitation,

$$u''(\tau) + \varepsilon\lambda u'(\tau) + u(\tau) + \varepsilon u^3(\tau) = \varepsilon A \cos[(1+\varepsilon\sigma)\tau] \tag{6}$$

where $\varepsilon A$ is the excitation amplitude, and $\sigma$ a frequency detuning parameter denoting the closeness of the excitation frequency to the linearized natural frequency, $\omega_n = 1$, of the Duffing oscillator. Equation (6) can be approximately solved by applying the multi-scale method [30], with steady-state solution approximately expressed as,

$$u(\tau) = a \cos[(1+\varepsilon\sigma)\tau - \gamma] + O(\varepsilon) \tag{7}$$

where the steady-state amplitude $a$ and phase $\gamma$ are determined by solving the slow-flow equations,

$$\lambda a = A \sin \gamma \tag{8a}$$

$$a\sigma - \frac{3}{8}a^3 = -\frac{A}{2}\cos \gamma \tag{8b}$$

Manipulating these equations, $\sigma$ can be determined as a function of $a$ by eliminating the phase $\gamma$, yielding the approximate expression for the nonlinear FRF (relating the amplitude $a$ to the frequency detuning $\sigma$ at steady state),

$$\sigma = \frac{3}{8}a^2 \pm \frac{1}{2}\left(\frac{A^2}{a^2} - \lambda^2\right)^{1/2} \tag{9}$$

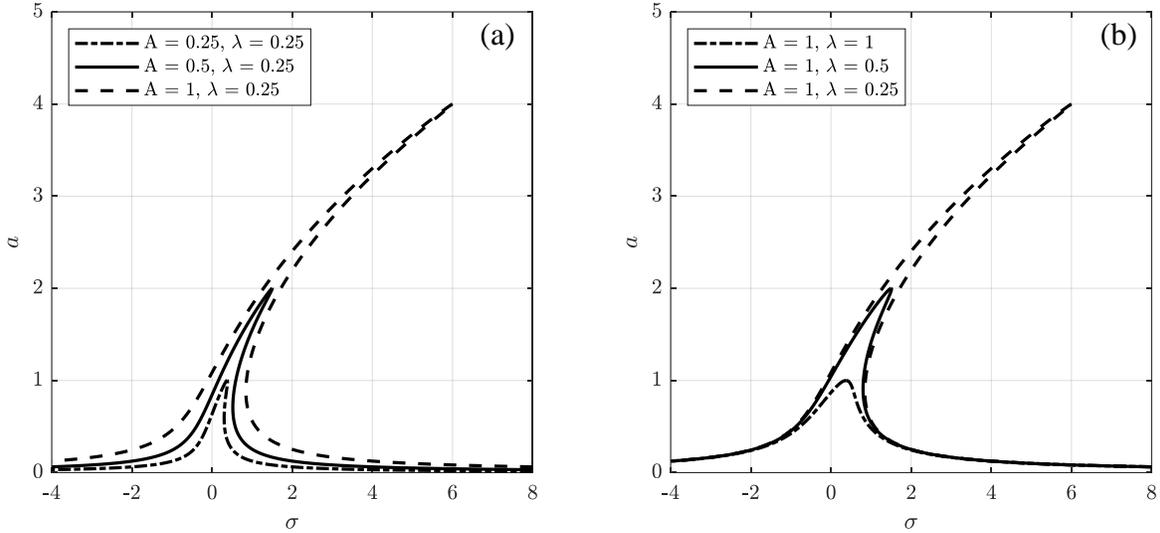

Figure 3. The effect of the forcing amplitude, $A$, and damping value, $\lambda$, on the frequency response function (FRF) of the damped Duffing oscillator (9): (a) FRF for fixed damping coefficient $\lambda = 0.25$ and varying excitation amplitude $A$; (b) FRF for fixed excitation amplitude $A = 1$ and varying damping coefficient $\lambda$.

Note that in the nonlinear case the FRF depends on the excitation amplitude, and the maximum value of the FRF is equal to $a_M = A/\lambda$. By definition, the half-power bandwidth is the difference of the side-band frequencies at which $a^2 = a_M^2/2$. Accordingly, substituting $a^2 = a_M^2/2$ into (9), the half-power bandwidth for the Duffing oscillator is found to be,

$$\Delta\omega_{-3\text{dB}} = \varepsilon\lambda \tag{10}$$

Eq. (10) reveals that the bandwidth of the damped Duffing oscillator, (5), is equal to the damping of the system and independent of the excitation amplitude; indeed, it is predicted to be the same as the bandwidth of the corresponding linearized system, cf. Fig. 2. This result, however, is clearly inconsistent with the nonlinear physics of the problem since one expects the half-power bandwidth of the nonlinear system to also be a function of the excitation amplitude, $A$. Moreover, this result is an approximation based on the dominant harmonic, so it ignores the very important effect on the bandwidth of higher harmonics which are typically generated in nonlinear systems. As such, it leaves an open question regarding the computation of the nonlinear bandwidth for other types of "hard" resonances, e.g., superharmonic and subharmonic resonances that are realized in the harmonically forced Duffing oscillator at different frequency ranges and for strong excitations. It is well known that the nonlinearly generated harmonics play the main role in the realization of these types of nonlinear resonance, which have no counterparts in linear theory. Lastly, like the classical half-power bandwidth definition for linear oscillators, the previous result is valid only for low-loss systems, i.e., weak dissipation, whereas there is an additional restrictive assumption of weak nonlinearity upon which the approximate solution (7-9) is based.

The previous example provides ample motivation for a new definition for the bandwidth of nonlinear oscillators. This is further emphasized by the observation that when considering a nonlinear oscillator, taking the Fourier transform of its free (decaying) response is not justified from a physical point of view since, as the frequency of the response varies with time or energy, it would be inconsistent with the basic assumption underlying the Fourier transform that the signal should be stationary. It follows that the Fourier-transformed free (decaying) responses cannot be used directly to compute the bandwidth of nonlinear oscillators. In addition, as the previous example shows, the FRF is likewise of no use in the nonlinear case. To overcome these limitations and provide a comprehensive new definition consistently valid for a general class of linear and nonlinear oscillators, we start by recognizing that the bandwidth of an oscillatory dynamical system is an indication of how fast (or slow) its energy is dissipated in the time domain, or, equivalently, how broadly the same energy is localized in the frequency domain. As stated earlier, the RMS bandwidth (3) of a signal represents its variance in the frequency domain [31]. According to the Fourier uncertainty principle, the variance of a signal in the frequency domain is inversely proportional to the variance of that signal in the time domain. This means that, for a signal with *high* dissipation (which results in localization of the signal in a *short* time), its RMS bandwidth must have a *large* value in the frequency domain that is (uniquely) in proportion with the time localization of the signal [31]. This consequently renders the RMS bandwidth a viable candidate for computing the bandwidth of (the energy of) general classes of oscillatory dynamical systems.

One possible difficulty with the proposed definition is the accurate computation of the energy of a SDOF linear/nonlinear oscillator, e.g., in an experimental system where the parameters have not been identified or are estimated with uncertainties. To overcome this potential issue, one might consider instead the envelope of its kinetic energy [32], which is approximately equal to its total instantaneous energy. Given that the kinetic energy is proportional to the velocity-squared of the SDOF oscillator (factored by an inertial term – e.g., its mass), it becomes more straightforward to measure the envelope of the velocity time series, $\langle v(t) \rangle$, a task that can be accomplished accurately (e.g., by fitting Akima spline curves to the series of the local maxima of the velocity [33]), and is always feasible in computations and experiments as long as measured velocity time series are available. Since the envelope of the velocity corresponding to the free response of a SDOF linear/nonlinear oscillator monotonically decays and does not possess time-varying frequencies, its Fourier transform, $V(\omega)$, can be computed and used to compute the bandwidth of the system according to the following expression (based on the RMS bandwidth definition):

$$\Delta\omega_{rms}^2 = \frac{4\int_0^\infty \omega^2 |V(\omega)|^4 d\omega}{\int_0^\infty |V(\omega)|^4 d\omega} \qquad (11)$$

In the following sections we will present a series of case studies to show the deficiency of the current half-power bandwidth for strongly nonlinear SDOF oscillators, compared to the extended bandwidth definition (11). Furthermore, we will demonstrate the efficacy of the extended bandwidth definition for SDOF oscillators with geometric nonlinearities and softening, hardening and bi-stable dynamics.

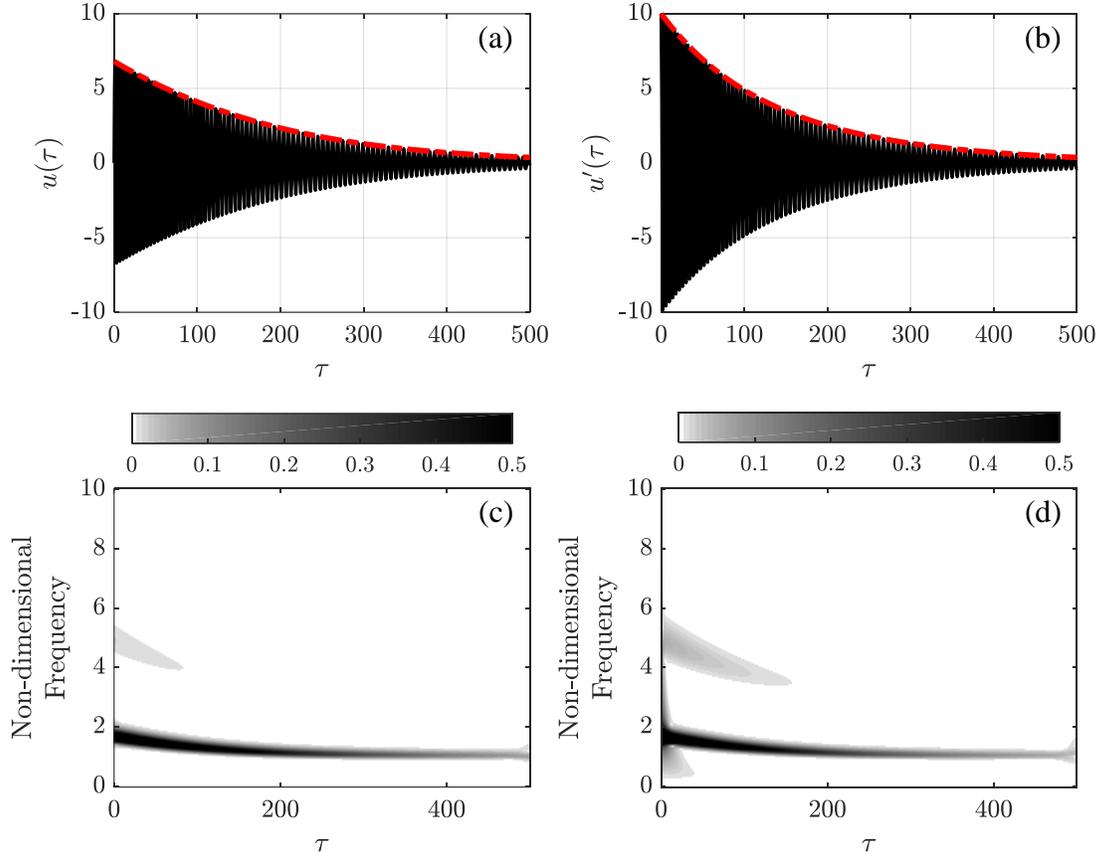

Figure 4. Free response of the damped Duffing oscillator with hardening nonlinearity (5), and initial conditions of $u(0) = 0$, $u'(0) = 10$, $\varepsilon = 0.05$ and $\lambda = 0.25$: (a) Displacement time series (black curve), $u(\tau)$ with corresponding envelope, $\langle u(\tau) \rangle$ (dash-dotted red curve), and (c) displacement wavelet spectrum; (b) velocity time series (black curve), $u'(\tau)$ with corresponding envelope, $\langle u'(\tau) \rangle$ (dash-dotted red curve), and (d) velocity wavelet spectrum.

## 3. Case studies

In this section, we employ the extended definition (11) to study the nonlinear bandwidth for two case studies. The first (and simpler) concerns the weakly damped and weakly nonlinear Duffing oscillator, (5), whereas the second addresses a dissipative SDOF strongly nonlinear oscillator which, depending on its geometry can possess hardening, softening or even bi-stability stiffness characteristics [34-37]. In particular, the strongly nonlinear dynamics of the second oscillator will demonstrate the capacity of (11) to model its energy-dependent bandwidth in different response regimes.

### 3.1 Damped Duffing oscillator with hardening nonlinearity

We begin by reconsidering the damped Duffing oscillator (5) with hardening nonlinearity. In Figure 4 the free decaying response of this system subject to initial conditions $u(0) = 0$ and $u'(0) = 10$, and parameters $\varepsilon = 0.05$, $\lambda = 0.25$ is depicted. Specifically, the displacement, $u(\tau)$, and velocity, $u'(\tau)$, together with their envelopes, $\langle u(\tau) \rangle$ and $\langle u'(\tau) \rangle$, respectively, are shown, based on which the bandwidth computation can be performed.

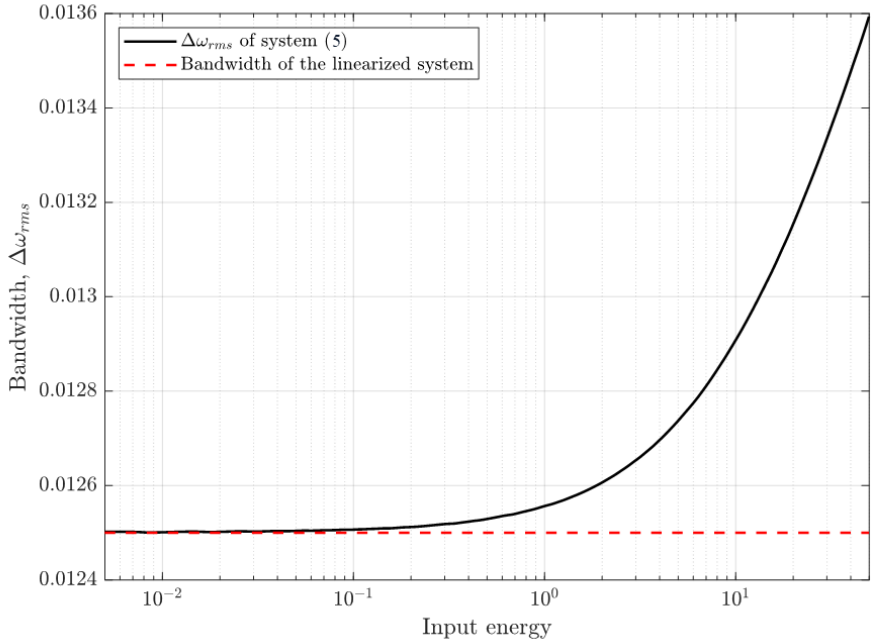

Figure 5. The nonlinear bandwidth, $\Delta\omega_{rms}$, of the damped Duffing oscillator (5) as a function of input energy $E_{in}$ (black curve); the dashed red line represents the bandwidth of the corresponding linearized system, where it holds that $\lim_{E_{in} \to 0} \Delta\omega_{rms} = \varepsilon\lambda = 0.0125$.

According to Figs. 4c and 4d, both $u(\tau)$ and $u'(\tau)$, contain high-frequency harmonics which complement the (stronger) fundamental harmonics. These harmonics will be implicitly reflected in the corresponding envelopes as well, which will be used to compute the bandwidth (11) at the specified input energy level, $E_{in} = {u'}^2(0)/2 + u^2(0)/2 + \varepsilon u^4(0)/4 = 50$. According to the extended bandwidth definition the velocity envelope of Fig. 4b is employed for this computation. In the following we proceed to compute the bandwidth of system (5) for varying $E_{in}$ through (11) with $V(\omega)$ being the Fourier transform of the velocity envelope $\langle u'(\tau) \rangle$, as depicted in Fig. 5.

Note that the nonlinear bandwidth, $\Delta\omega_{rms}$, of the hardening Duffing oscillator is a *monotonically increasing* function of its input energy. As stated earlier, the bandwidth of a signal is related to its decay rate in the time domain, which for a dynamical system quantifies its dissipative capacity. According to Fig. 5, as the energy input to the system increases, its capacity to dissipate energy increases as well. From a physical standpoint, this is expected since the damped Duffing oscillator nonlinearly scatters energy to higher frequencies through the generation of higher harmonics, as shown in Figs. 4c and 4d. In turn, this nonlinear frequency scattering yields accelerated energy dissipation compared to the linearized system (i.e., the linear weakly damped oscillator obtained in the limit of small input energy), since energy is more efficiently and rapidly dissipated with increasing frequency. Indeed, in the limit of small input energy levels, the nonlinear effects and the term $\varepsilon u^3$ in (5) become negligibly small, and the Duffing oscillator (5) degenerates to a SDOF linear oscillator similar to (1) with $f(t) = 0$, $\omega_0 = 1$ and $2\zeta = \varepsilon\lambda$; as mentioned earlier, the classical half-power bandwidth of this linear, lightly damped system is equal to its damping coefficient $\varepsilon\lambda$. Hence, the result in Fig. 5 confirms that the extended bandwidth definition (11) correctly converges to the classical half-power bandwidth in the limit of small energies, $\lim_{E_{in}\to 0} \Delta\omega_{rms} = \Delta\omega_{-3\text{dB}} = \varepsilon\lambda = 0.0125$.

This first case study shows that, unlike the classical half-power bandwidth, the extended definition (11) is valid for nonlinear SDOF oscillators, as it considers the harmonic generation caused by the nonlinearity. As such, it is energy-dependent, is not limited by the low-loss (weak dissipation) assumption, and does not require a FRF computation, which for nonlinear systems is problematic; rather, the extended bandwidth computation is based solely on the measurement of the free response (velocity) of the system, which in most cases can be conveniently performed experimentally. Lastly, the extended bandwidth (11) correctly converges to the classical half-power bandwidth in the limit of negligible small nonlinear effects, which indicates that is a generalization for the nonlinear case of that linear concept.

**3.2 SDOF geometrically nonlinear oscillator with hardening, softening and bistability**

The second case study focuses on a SDOF nonlinear oscillator with more complicated dynamics when compared to the Duffing oscillator. This oscillator was introduced in earlier work [34-37] and is configured as shown in Figure 6; that is, as a lumped mass grounded by a linear spring-viscous damper pair and by an initially inclined linear spring. It is assumed that this system is restricted to oscillate in the horizontal direction and that no external excitation exists. Assuming zero initial displacement and arbitrary initial velocity $v_0$, the governing nonlinear equation of motion is given by,

$$m\ddot{x}(t) + d_l\dot{x}(t) + k_l x(t) + k_i y(t)\left[1 - \frac{\sec\phi_0}{\sqrt{L^2 + y^2(t)}}\right] = 0$$

$$x(0) = 0, \dot{x}(0) = v_0 \tag{12}$$

where $y(t) = x(t) + L\tan\phi_0$, $\phi_0$ denotes the initial angle of inclination, overdot implies differentiation with respect to time $t$, and the remaining system parameters are defined in Fig. 6. Normalizing the system (12) one obtains,

$$u''(\tau) + \lambda_l u'(\tau) + u(\tau) + \beta\hat{y}(\tau)\left[1 - \frac{\sec\phi_0}{\sqrt{1+\hat{y}^2(\tau)}}\right] = 0$$

$$u(0) = 0, u'(0) = w_0 \qquad (13)$$

where $\tau = t(k_l/m)^{1/2}$, $\lambda_l = d_l(k_l m)^{-1/2}$, $\beta = k_i/k_l$, $u = x/L$, and $\hat{y} = y/L$. One notes that the inclined linear spring introduces strong geometric stiffness nonlinearity to this system.

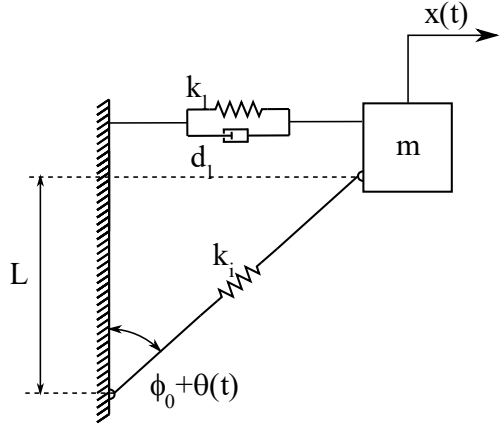

Figure 6. The geometrically nonlinear oscillator grounded by a linear spring-viscous damper pair and an inclined linear spring with angle of inclination $\phi_0$.

Considering more closely the nonlinear stiffness term in (13), $u + \beta\hat{y}(1 - \sec\phi_0/\sqrt{1+\hat{y}^2})$, it can be shown that depending on the stiffness ratio $\beta$ and the initial angle of inclination $\phi_0$ *the resulting nonlinear dynamics can be tuned to be either softening or hardening*. Even more interesting is that, depending on $\beta$, there is a critical value of $\phi_0 = \phi_{cr}(\beta) = \cos^{-1}\{(\beta - 1)/[(\beta + 2)\sqrt{\beta(\beta+2)/(\beta^2-1)}]\}$ where a saddle node bifurcation occurs generating *bistability* in the dynamics. Specifically, for $\phi_0 < \phi_{cr}(\beta)$ there exists only a single trivial stable equilibrium position, which persists for all values of $\beta$, whereas for $\phi_0 > \phi_{cr}(\beta)$ two additional non-trivial equilibria are generated by the bifurcation; one of these new non-trivial equilibria is stable and the other unstable. Hence, the bifurcation qualitatively changes the free dynamics of this nonlinear oscillator for small or large angles of inclination, as shown by the simulations of Figure 7. These depict four different response regimes of (13) for $\lambda_l = 0.05$, $\beta = 200$, different normalized initial velocities $w_0$, and initial inclination angles below or above the critical angle $\phi_{cr}(200)\sim 11.4°$.

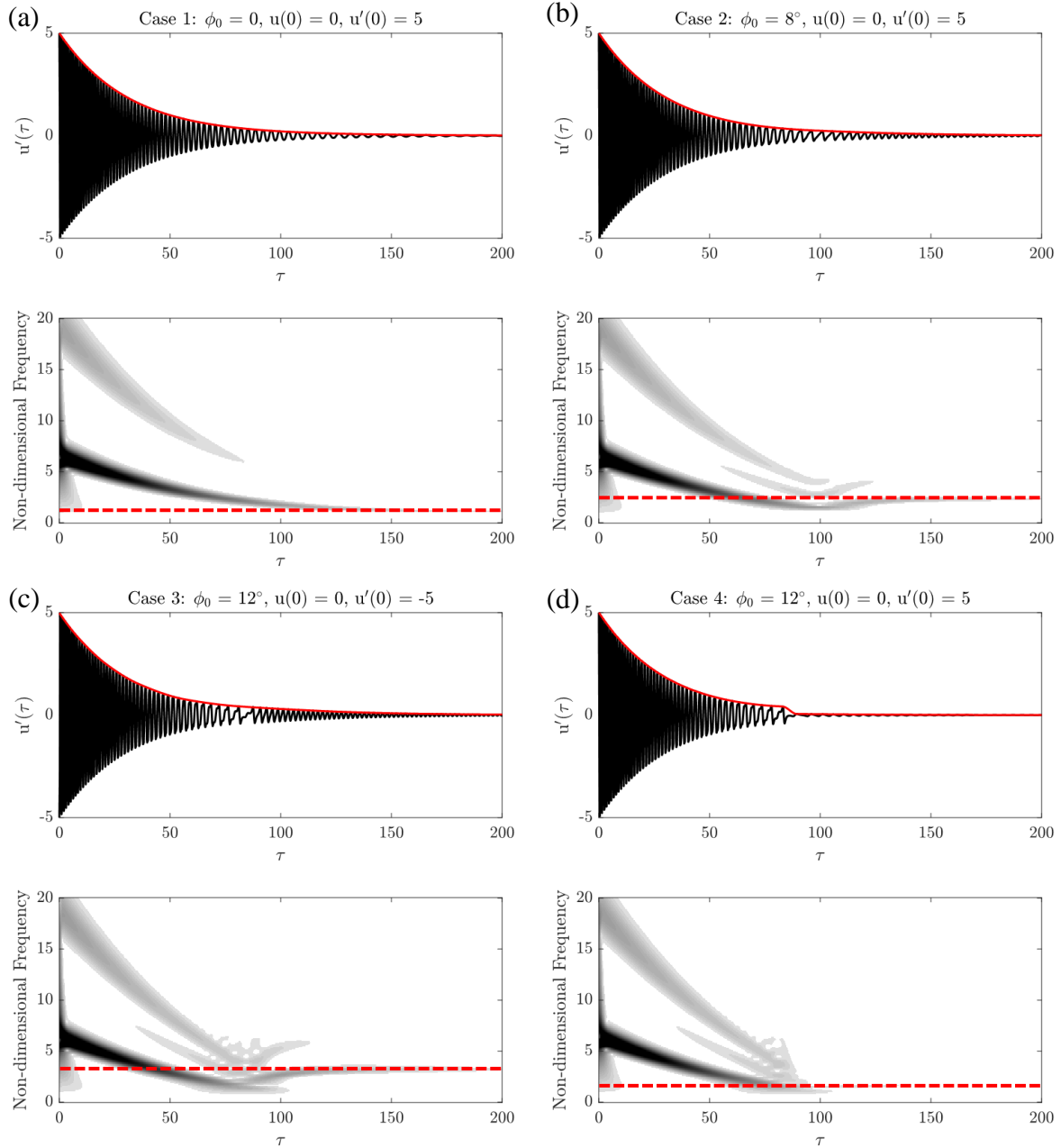

Figure 7. Four different dynamical regimes of oscillator (13) are shown, depending on the initial angle of inclination and initial velocity: Displacement and velocity responses (red lines indicate envelopes) with their corresponding wavelet transforms for (a) $\phi_0 = 0$, $w_0 = 5$ (hardening response), (b) $\phi_0 = 8°$, $w_0 = 5$ (hardening-softening response), (c) $\phi_0 = 12°$, $w_0 = -5$ (bistable response), and (d) $\phi_0 = 12°$, $w_0 = 5$ (bistable response); dashed red line indicates the linearized natural frequency of the system in the limit of very small energy.

Starting with Fig. 7a corresponding to $\phi_0 = 0$ and $w_0 = 5$, the geometric nonlinearity is of hardening type. This is confirmed by the monotonic decay of the oscillation frequency as the energy of the system decreases, i.e., time lapses – cf. the wavelet transform contour in Fig. 7a. On the other hand, the response shown in Fig. 7b, corresponding to $\phi_0 = 8°$ and $w_0 = 5$, possesses both hardening (in the early, highly energetic regime of the dynamics) and softening nonlinearities (in the later regime where the energy is smaller). Indeed, the hardening nonlinearity dominates the dynamics from $\tau = 0$ to $\tau = 100$, as shown by the decreasing frequency with decreasing energy in that response regime. However, for $\tau > 100$, the frequency of oscillation increases with decreasing energy, which indicates softening nonlinearity.

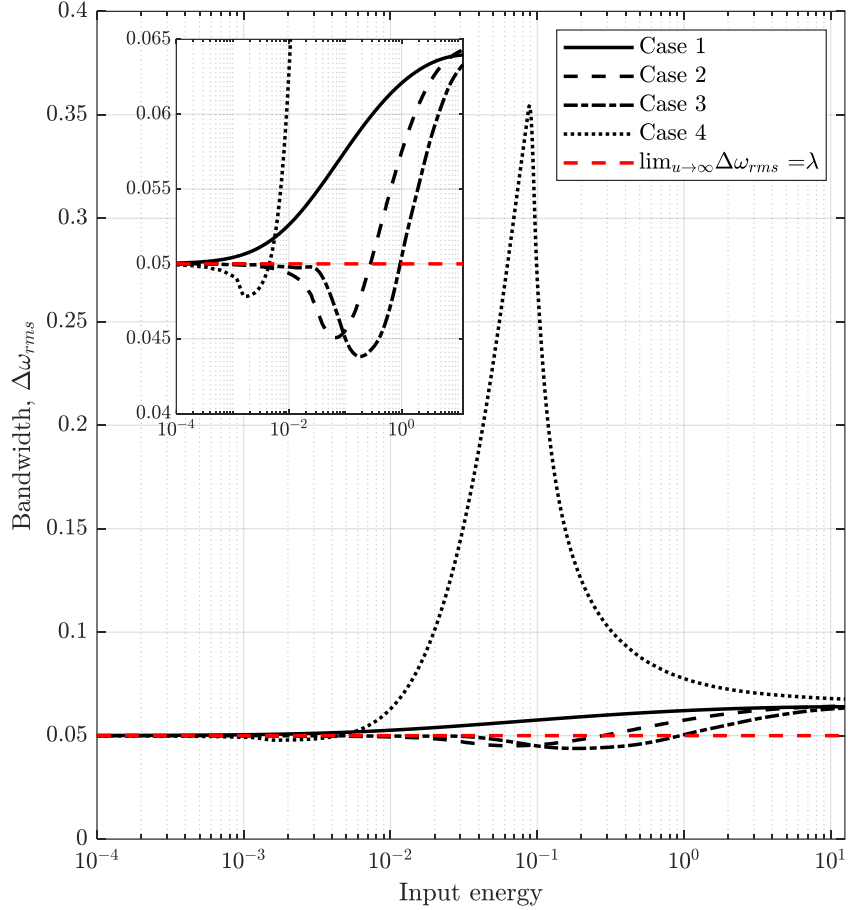

Figure 8. The energy-dependent bandwidth (11) of the geometrically nonlinear oscillator (13) for varying input energy and $\phi_0 = 0$ (case 1 – solid curve), $\phi_0 = 8°$ (case 2 – dashed curve), $\phi_0 = 12°$ when the trivial equilibrium is reached (case 3 – dash-dotted curve), and $\phi_0 = 12°$ when the non-trivial equilibrium is reached (case 4 – dotted curve); the dashed red line shows the classical half-power bandwidth of the linearized oscillator as $\lim_{u \to 0} \Delta \omega_{rms} = \lambda = 0.05$, and the inset shows a detail of the plots for input energy in the range $0.04$ to $0.065$.

Further increasing the initial inclination angle beyond the critical value of $\phi_{cr} \sim 11.4°$ yields bistability in the dynamics. This response regime is considered in Figures 7c and 7d for $\phi_0 = 12°$, i.e., just above the bifurcation point, and initial velocities $w_0 = -5$ and $w_0 = 5$, respectively. Considering the response in Fig. 7c, the oscillator settles in the trivial fixed point, $\lim_{\tau \to \infty} u = 0$, whereas for a different initial condition (cf. Fig. 7d) it settles into the non-trivial stable equilibrium which is generated by the bistability. This is deduced from the sudden decrease in the amplitude of the velocity-time signal at $\tau \sim 90$ in the latter case, and is attributable to the attraction of the dynamics by the non-trivial stable fixed point, so that $\lim_{\tau \to \infty} u \neq 0$; in this case a portion of the total energy of the system is permanently transformed into potential energy, thereby causing both the corresponding kinetic energy and velocity amplitude to suddenly decrease. Viewed from a different perspective, if we consider the kinetic energy to be an indicator of effective output of the system, then the sudden drop of the kinetic energy in Fig. 7d portends the sudden increase of the dissipative capacity of the oscillator due to bistability.

Based on the time series of Figure 7 one proceeds to compute the nonlinear bandwidth (11) of the oscillator (13) in the previous response regimes. To this end, it is of interest to compute the bandwidth for the entire input energy range corresponding to initial velocities $w_0 \in [0.01,5]$, and the different initial inclination angles considered previously, i.e., smaller or larger than the critical bifurcation value. In the following we refer to $\phi_0 = 0$ as case 1, $\phi_0 = 8°$ as case 2, $\phi_0 = 12°$ as case 3 when the dynamics settles into the trivial equilibrium, and, lastly, $\phi_0 = 12°$ as case 4 when the dynamics settles into the non-trivial equilibrium. In Figure 8 the energy-dependent bandwidth of (13) is depicted for cases 1 to 4 and compared to the classical half-power bandwidth of the linearized system obtained at the low energy limit, i.e., $\lim_{u \to 0} \Delta\omega_{rms} = \lambda = 0.05$. Starting with case 1 ($\phi_0 = 0$) the bandwidth monotonically increases with increasing energy, above that of the linearized system. To give an indication, at an input energy level of $\sim 10^1$ the hardening nonlinearity increases the bandwidth by approximately 30% compared to the linearized case. From a physical perspective, and like the Duffing oscillator considered in the previous section, the increase in the bandwidth (and therefore in the dissipative capacity of the system) is due to the hardening stiffness nonlinearity which generates high frequency harmonics by nonlinearly scattering energy in the frequency domain – cf. Fig. 7a, which, in turn, yields an increased rate of energy dissipation.

Proceeding to case 2 ($\phi_0 = 8°$) the dynamics change significantly compared to case 1, since, depending on the level of its instantaneous energy, the oscillator possesses either softening (at large energy) or hardening (at low energy) stiffness nonlinearity. This is directly reflected in the energy dependent bandwidth of the system, since for sufficiently high input energy – where the nonlinearity is hardening, the bandwidth increases (approximately by 30% for input energy of $\sim 10^1$) compared to the linearized case; the physical interpretation is like that of case 1. This happens for the highly energetic regime of the transient dynamics. However, for slightly lower input energy level, the dynamics starts with and is dominated by softening stiffness nonlinearity, hence, the bandwidth decreases below the linearized bandwidth $\lim_{u \to 0} \Delta\omega_{rms} = \lambda = 0.05$ (e.g., it becomes approximately 10% less at input energy $\sim 7 \times 10^{-2}$). This can be explained by examining the frequency of oscillation during this softening phase of the dynamics, which becomes less than the linearized natural frequency, i.e., $\omega_n^l = \sqrt{1 + \beta \sin^2 \phi_0}$ [36,37]. Thus, the oscillations of the system become slower, resulting, in turn, in a slower dissipation rate of energy by the viscous damper. Accordingly, when the transient dynamics of the oscillator is dominated by the softening nonlinearity, the dissipative capacity of the oscillator (13) deteriorates, and its bandwidth decreases.

Cases 3 and 4 are in the bistability regime of the dynamics, corresponding to the same inclination angle, $\phi_0 = 12°$, but different initial conditions which result in attraction by different stable equilibrium positions of the oscillator [34,37]. Specifically, in case 3, the transient dynamics settles into the trivial stable equilibrium (cf. Fig. 7c), and the corresponding bandwidth has an energy dependence which is qualitatively like case 2. A drastically different result, however, is obtained for case 4 (cf. Figs. 7d and 8), where the transient dynamics can be divided into four distinct regimes. Starting from the highly energetic regime ($0 < \tau < 70$), the dynamics is of the hardening type and similar to case 1. For high enough input energy levels for which the dynamics of the system is dominated by the hardening nonlinearity (where the attraction of the oscillator to the non-trivial fixed-point occur at late times), the bandwidth is greater than its linearized limit. Next is the regime of transient attraction to the stable non-trivial equilibrium point ($70 < \tau < 90$) - cf. Fig. 7d. For input energy levels where the dynamics is dominated by the regime of attraction to the stable non-trivial fixed-point, i.e., non-dimensional input energy in the range of $10^{-2}$ and $10^0$, the bandwidth increases drastically (by as much as $600\%$ at an input energy level of $\sim 0.09$). This is expected since during this regime a portion of the kinetic energy of the oscillator permanently transforms to potential energy as the dynamics is attracted to the nontrivial equilibrium. Accordingly, this is accompanied by a sudden decrease in velocity, i.e., much higher effective energy dissipation, and, in turn, much higher bandwidth. Hence, when bistability is triggered and the dynamics of the oscillator is attracted to the stable non-trivial equilibrium, the bandwidth becomes much greater compared to the other cases. In the next regime, as energy decreases even further, the oscillator performs decaying oscillations about the non-trivial equilibrium, and the dynamics becomes softening. For initial input energy levels corresponding to this regime of dynamics, i.e., input energy levels for which this regime of dynamics occur at early times, the bandwidth below the linearized limit. Lastly, in the final, very low energy regime, the transient dynamics become linearized. This translates to low input energy levels for which the bandwidth approaches asymptotically the linearized half-power bandwidth, $\lim_{u \to 0} \Delta\omega_{rms} = \lambda = 0.05$, from below.

Apart from validating the extended bandwidth definition (11) from a physical point of view, the previous examples demonstrate its efficacy for even strongly nonlinear SDOF oscillators, where the traditional linear-based definition clearly does not hold. In addition, the results show that depending on the nature of the nonlinear dynamics, i.e., hardening, softening, or bi-stable, and the instantaneous energy level of the oscillation, the nonlinear bandwidth can become larger (even drastically larger) and/or smaller than the linearized half-power bandwidth obtained in the limit of small energy. Even more interesting is the demonstration that the bandwidth of a nonlinear oscillator can vary drastically both above and below the linearized limit during the same transient oscillation, provided that different transitions occur between qualitatively different dynamical regimes during the motion. A more detailed study of the nonlinear transient dynamics will be undertaken in the next section, to study and explain how nonlinearity and in particular nonlinear scattering of energy in the frequency domain during the transient response, influences the bandwidth of a nonlinear SDOF dissipative oscillator.

## 4. Bandwidth and nonlinear harmonic generation

The main feature of nonlinearity is the generation of harmonics, that is, the scattering of the input energy in the frequency domain (and the wavenumber domain in the case of acoustics). It is natural, therefore, to seek the link between nonlinear higher or lower harmonic generation and the bandwidth in nonlinear oscillators. To this end, the transient response of a nonlinear dissipative oscillator is expressed as a series superposition of harmonics in the time domain,

$$u(\tau) \approx \sum_{i=1}^{N} u_i(\tau) \tag{14}$$

where $u_i(\tau)$ denotes the $i$-th harmonic component (with $u_1(\tau)$ denoting the fundamental harmonic), and where the series is truncated beyond the $N$-th harmonic. As shown in [38] through the use of inverse wavelet transforms, the harmonic decomposition (14) is always possible in nonlinear responses. At this point one recalls that the extended bandwidth (11) is based on the *envelope* of the velocity signal, $\langle u'(\tau) \rangle$, which, through the series (14) can be expressed in terms of the response harmonics as:

$$\langle u'(\tau) \rangle \approx \langle \sum_{i=1}^{N} u_i'(\tau) \rangle = \sum_{i=1}^{N} \alpha_i \langle u_i'(\tau) \rangle \tag{15}$$

The coefficients $\alpha_i$ account for the phase differences between the harmonic velocities $u_i'$. Setting $\alpha_1 = 1$ and evaluating the other coefficients by minimizing the following loss-function given as

$$\mathcal{L} = (\langle u' \rangle - \sum_{i=1}^{N} \alpha_i \langle u_i' \rangle)^2 + \sum_{i=1}^{N} \alpha_i^2 \tag{16}$$

At this point, one employs expression (15) to express the bandwidth formula (11) in terms of the harmonics (14) as,

$$\Delta \omega_{rms}^2 = \frac{\sum_{i=1}^{N} \alpha_i^4 E_{v_i}^2 \Delta \omega_i^2}{E_v^2} + \frac{\int_{-\infty}^{\infty} \omega^2 [V^4(\omega) - \sum_{i=1}^{N} \alpha_i^4 V_i^4(\omega)] d\omega}{E_v^2} \tag{17}$$

where $V(\omega) = \mathcal{F}\{\langle u'(\tau) \rangle\}$, $V_i(\omega) = \mathcal{F}\{\langle u_i'(\tau) \rangle\}$, $E_v^2 = \int_0^\infty |V(\omega)|^4 d\omega$, $E_{v_i}^2 = \int_0^\infty |V_i(\omega)|^4 d\omega$, and $\mathcal{F}\{\bullet\}$ represents the Fourier transform operator. Moreover, the (energy-dependent) bandwidth of the $i$-th harmonic is computed according to expression (11) as,

$$\Delta \omega_i = 2 \sqrt{\frac{\int_0^\infty \omega^2 |V_i(\omega)|^4 d\omega}{\int_0^\infty |V_i(\omega)|^4 d\omega}} \tag{18}$$

Expression (17) directly relates the extended bandwidth (11) to nonlinear harmonic generation. Clearly, if there are no nonlinear harmonics generated (i.e., in the absence of nonlinearity) and only the fundamental harmonic exists, expression (17) recovers the traditional half-power bandwidth. Hence, (17) enables the detailed exploration of the contributions of the nonlinear harmonics to the bandwidth (11).

Note that the first term in (17) represents the contribution of the weighted summation of the individual harmonics (weighted by $E_{v_i}^2/E_v^2$) to the bandwidth, so it quantifies the contribution of each individual harmonic. In contrast, the second term in (17) quantifies the effects of inter-harmonic interactions (cross-harmonic product terms), i.e., the contributions to the bandwidth (11) of inter-harmonic energy exchanges. Therefore, the important expression (17) provides a fundamental tool for exploring how nonlinear harmonic generation, and energy exchanges between harmonics affect the bandwidth. Viewed in another context, the expression (17) enables the direct study of *nonlinear energy scattering in the frequency domain* and its effect on the nonlinear bandwidth; hence, it provides a powerful new tool to study in detail the nonlinear physics and the strength of the nonlinear effects, as evidenced by the capacity of the nonlinearity to intensely scatter energy in the frequency domain.

As an application the oscillator (13) is reconsidered for $\phi_0 = 0$ (case 1 – hardening nonlinearity) and $\phi_0 = 8°$ (case 2 – hardening and softening nonlinearity) with common initial velocity $w_0 = 10$, in order to correlate the corresponding nonlinear harmonic generation to the energy dependent bandwidth as depicted in Fig. 8. Staring with the case of hardening nonlinearity and the plots of Figure 9, one notes that there are two dominant harmonics (first and third) in the response (there exists a countable infinity of higher harmonics of smaller amplitude, the effects of which are negligible); hence, one truncates the series (15) and (17) at $N = 3$. However, before the contributions of the individual harmonics (and their interactions) can be evaluated, it is necessary to extract the velocity time series associated with each harmonic, separately. To do so, one employs the recently developed inverse continuous wavelet transform (ICWT) method for mode decomposition developed by Mojahed et al. [38], yielding the harmonic decomposition in the time and wavelet domains depicted in Figure 10.

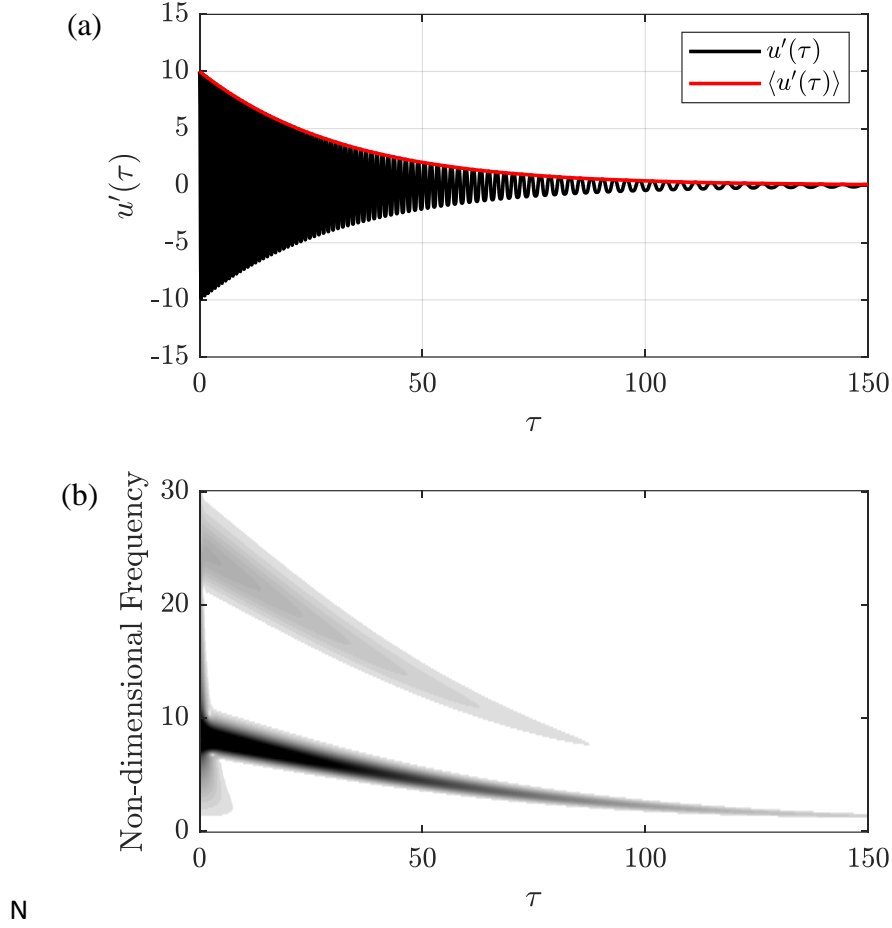

Figure 9. Response of oscillator (13) for $\phi_0 = 0$ and $w_0 = 10$: (a) Velocity time series and envelope, and (b) corresponding wavelet transform; this is a case of hardening stiffness nonlinearity.

Specifically, the plots of Figs. 10b,c and 10d,e depict the decomposed harmonics, $u'_1(\tau)$ and $u'_3(\tau)$, in terms of which the original velocity signal, $u'(\tau)$, can be reconstructed by direct superposition, $u'(\tau) \approx u'_1(\tau) + u'_3(\tau)$ with an accuracy of $R^2 = 0.9992$, where $R^2$ is the coefficient of determination [38]. The Fourier transforms of the envelopes $\langle u'_1(\tau) \rangle$ and $\langle u'_3(\tau) \rangle$ of the decomposed velocity harmonics can now be substituted into (17) to determine the contribution of these harmonics to the bandwidth of (13); the multiplicative coefficients are computed by minimizing (16), yielding $\alpha_1 = 1$ and $\alpha_2 = -1$. It follows that in this case the bandwidth is approximated by the following truncated relation,

$$\Delta \omega_{rms}^2 \approx \frac{\Delta \omega_1^2 \int_0^\infty |V_1(\omega)|^4 d\omega}{E_v^2} + \frac{\Delta \omega_3^2 \int_0^\infty |V_3(\omega)|^4 d\omega}{E_v^2} + \frac{\int_{-\infty}^\infty \omega^2 [V^4(\omega) - V_1^4(\omega) - V_3^4(\omega)] d\omega}{E_v^2} \quad (19)$$

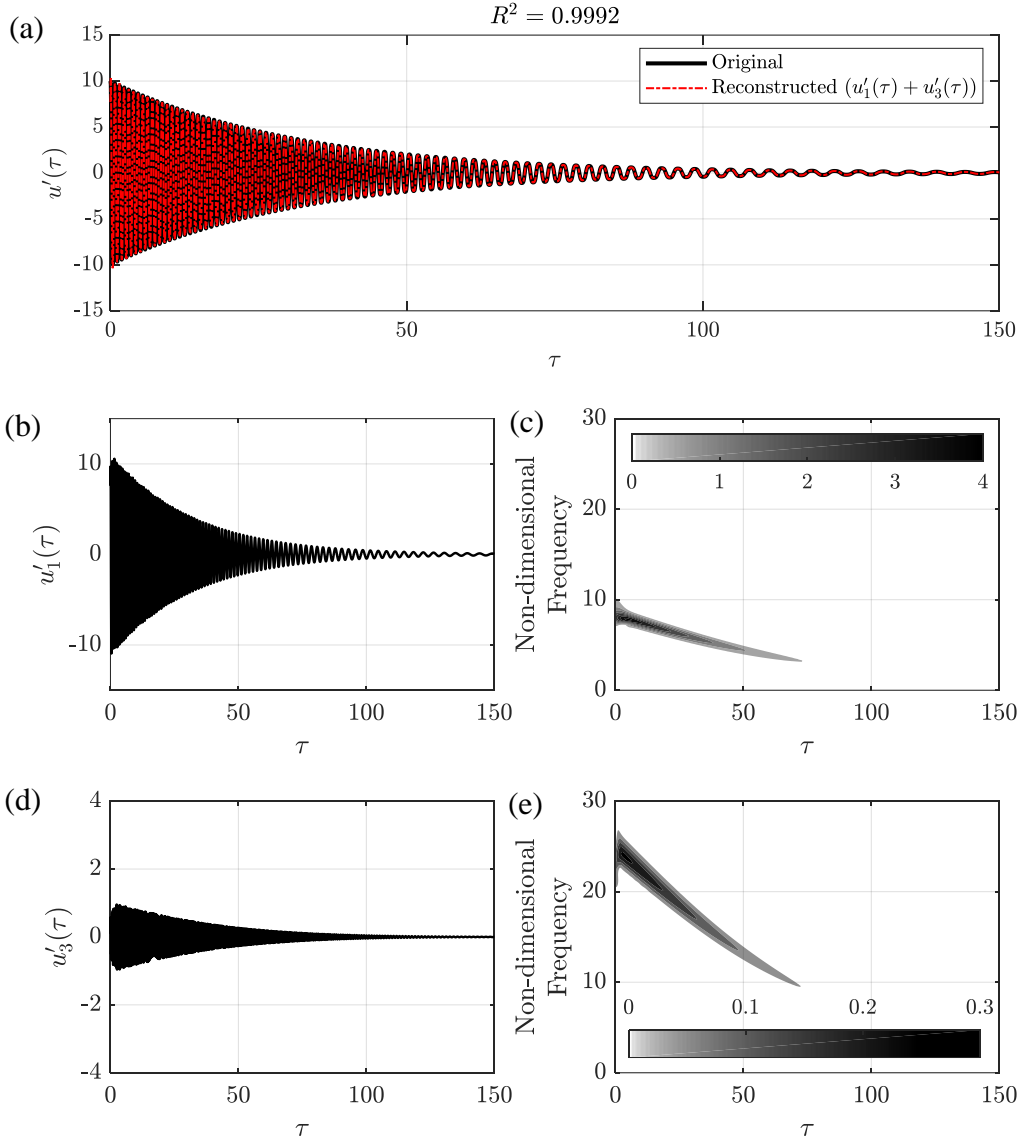

Figure 10. Response of oscillator (13) for $\phi_0 = 0$ and $w_0 = 10$ – hardening stiffness nonlinearity: (a) Velocity time series $u'(\tau)$ and reconstruction $u'_1(\tau) + u'_3(\tau)$; decomposed (through inverse wavelet transform [38]) of the (b,c) fundamental harmonic $u'_1$, and (d,e) third harmonic $u'_3$ in the time and wavelet domains, respectively.

where the two leading terms represent the individual contributions of the first and third harmonics, and the third term is an interaction term quantifying the contribution to the bandwidth of energy exchanges between the first and third harmonics. As mentioned previously, higher harmonics are neglected since their contributions are negligible.

In Figure 11 the energy-dependent $\Delta\omega^2_{rms}$ as well as the contributions of the individual terms in (19) are depicted. As mentioned, the bandwidth of a nonlinear oscillator is a measure of its overall dissipative capacity for varying input energy. From Fig. 11 the bandwidth contribution of the fundamental harmonic is greater than the total bandwidth. This means that the fundamental harmonic loses energy at a faster rate compared to the overall transient dynamics. The third harmonic, on the other hand, does not contribute *individually* to the bandwidth; however, it does contribute to it through *its nonlinear interaction with the fundamental harmonic*, i.e., through the energy exchange with the fundamental harmonic. As stated earlier, the fundamental harmonic loses its energy at a faster rate compared to the original system; however, this energy is not entirely dissipated by the first harmonic, since a part of it is transferred to the higher (third) harmonic. This explains the negative contribution to the bandwidth of the inter-harmonic interaction term, indicating that the third harmonic continuously gains energy with the first harmonic. This increases the overall rate of energy dissipation by the nonlinear oscillator as energy is dissipated faster at higher frequencies, which, in turn, increases the bandwidth. Note that, as the input energy tends to zero, the effects of nonlinearity diminish, and the dynamics become linearized. This is also evident by the nonlinear inter-harmonic interaction term in Fig. 11, which tends to zero as the dynamics become linearized and harmonic generation is eliminated; then, the entire contribution to the bandwidth comes from the fundamental harmonic (as it should).

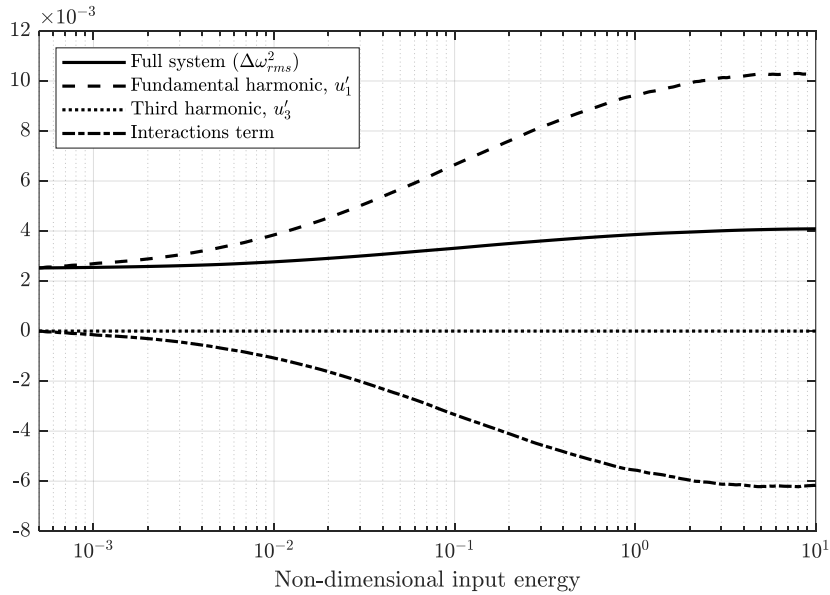

Figure 11. Contributions to the energy-dependent bandwidth of the three terms in (19): Contribution of the fundamental harmonic is shown by the dashed curve, of the third harmonic by the dotted curve and of the inter-harmonic interactions by the dash-dotted curve; the resulting bandwidth $\Delta\omega_{rms}^2$ is depicted by the solid curve (cf. also case 1 in Fig. 8).

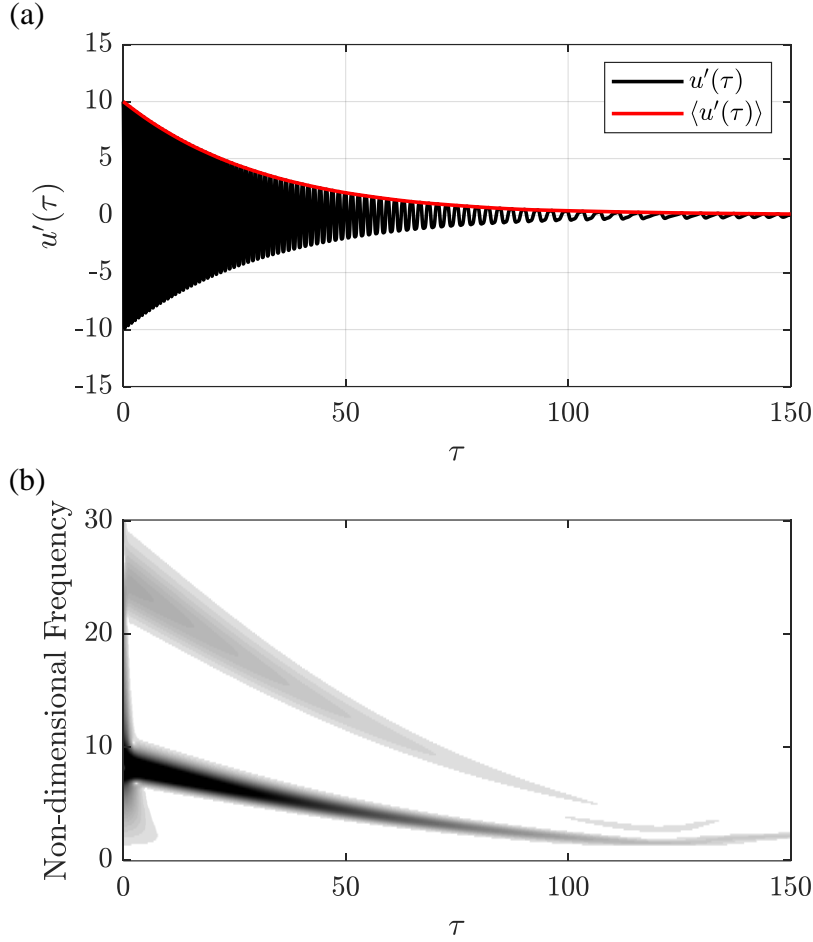

Figure 12. Response of oscillator (13) for $\phi_0 = 8°$ and $w_0 = 10$: (a) Velocity time series and envelop, and (b) corresponding wavelet transform; this is a case of combined hardening and softening stiffness nonlinearity.

Proceeding to the next case, we consider oscillator (13) with $\phi_0 = 8°$ and $w_0 = 10$, which possesses both hardening nonlinearity (in the early, highly energetic regime of the dynamics) and softening nonlinearity (in the later, lower energetic regime). Figure 12 depicts the corresponding normalized velocity signal, the key difference here with the case of purely hardening nonlinearity of Fig. 9 being the generation of a second harmonic in the softening regime of the transient dynamics. Similar to the previous case, it is necessary to decompose the three leading harmonic components of the response using the inverse continuous wavelet transform approach [38]. The decomposed harmonics are depicted in Figure 13, with $\alpha_1 = 1, \alpha_2 = 0.5696$, and $\alpha_3 = -1.0158$. As an indication of the accuracy of the harmonic decomposition, the velocity time series $u'(\tau)$ is reconstructed in terms of the superposition $u'_1(\tau) + u'_2(\tau) + u'_3(\tau)$ of the three leading harmonics with an accuracy of $R^2 = 0.9991$. Moreover, the wavelet transforms depicted in Figs. 13c,e,g further confirm the monochromatic nature of the decomposed harmonics $u'_i(\tau), i = 1,2,3$. With this information one can compute the approximate bandwidth through the truncated expression,

$$\Delta\omega_{rms}^2 \approx \frac{\Delta\omega_1^2 \int_0^\infty |V_1(\omega)|^4 d\omega}{E_v^2} + \frac{0.1053\Delta\omega_2^2 \int_0^\infty |V_2(\omega)|^4 d\omega}{E_v^2} + \frac{1.065\Delta\omega_3^2 \int_0^\infty |V_3(\omega)|^4 d\omega}{E_v^2} +$$
$$\frac{\int_{-\infty}^\infty \omega^2[V^4(\omega) - V_1^4(\omega) - 0.1053\, V_2^4(\omega) - 1.065\, V_3^4(\omega)] d\omega}{E_v^2} \tag{20}$$

As in the previous case, the first three terms in (20) represent contributions to the bandwidth of the individual harmonics, whereas the last term captures the contributions of inter-harmonic interactions.

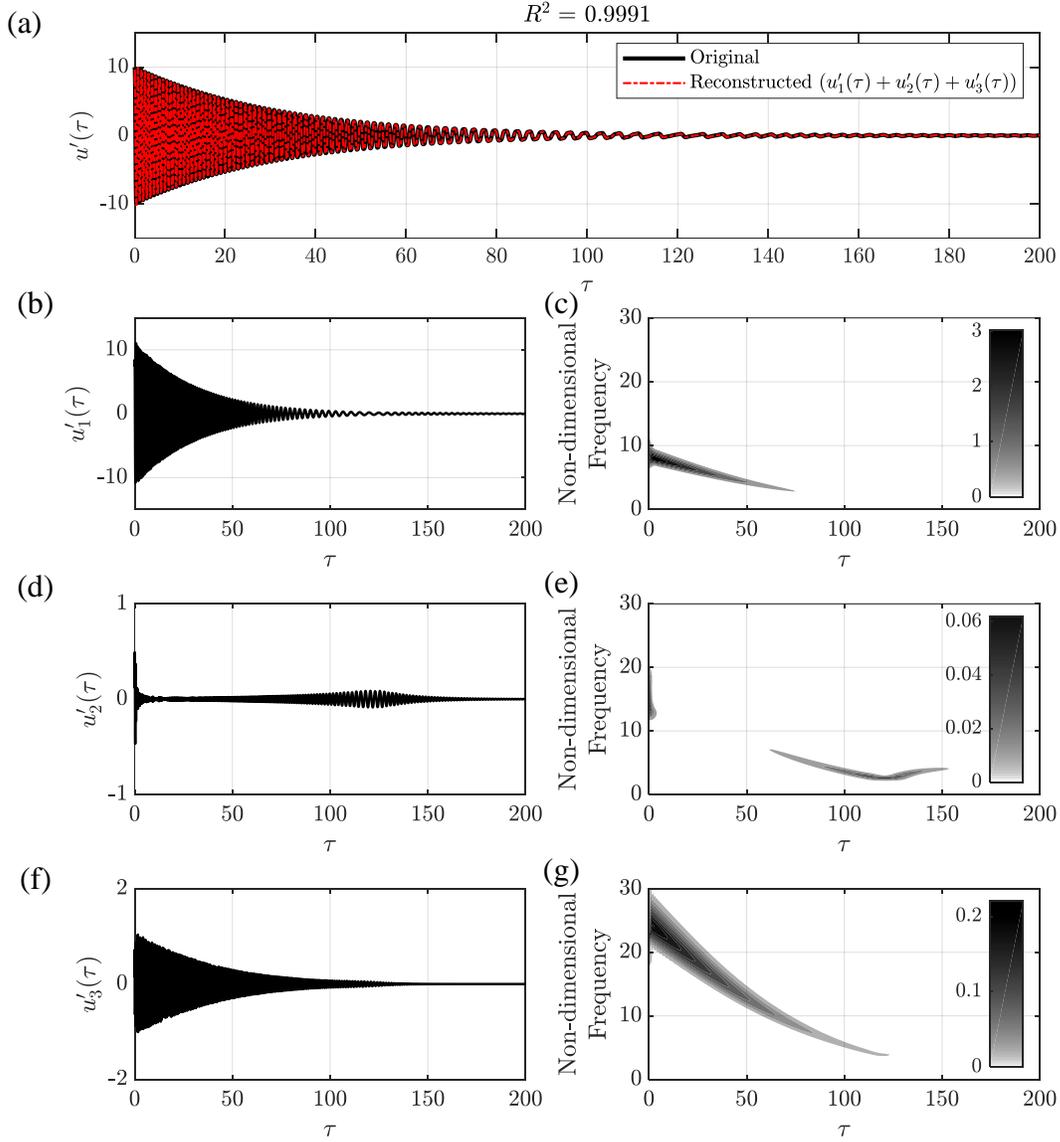

Figure 13. Response of oscillator (13) for $\phi_0 = 8°$ and $w_0 = 10$ – combined hardening and softening stiffness nonlinearity: (a) Velocity time series $u'(\tau)$ and reconstruction $u_1'(\tau) + u_2'(\tau) + u_3'(\tau)$; decomposed (through inverse wavelet transform [38]) of the (b,c) fundamental harmonic $u_1'$, (d,e) second harmonic $u_2'$, and (f,g) third harmonic $u_3'$ in the time and wavelet domains, respectively.

Figure 14 depicts the energy-dependent bandwidth contribution of each term in (20) for this system. Immediately, certain similarities and differences between these results and those for the purely hardening case depicted in Fig. 11 can be identified. For instance, in both cases the fundamental harmonic and the inter-harmonic interaction terms contribute mostly to the energy-dependence of the bandwidth, while the individual contributions of the higher harmonics are negligible. Moreover, as in the previous case the higher harmonics manifest themselves through their interactions (targeted energy transfers) with the fundamental harmonic as well as the inter-harmonic interactions between them. In contrast with the previous case, the inter-harmonic interaction term contributes negatively to the bandwidth only during the (highly energetic) hardening regime of the transient dynamics, whereas at intermediate-to-lower energy regimes the inter-harmonic interaction term becomes positive signifying an interesting reversing trend in the nonlinear dynamics.

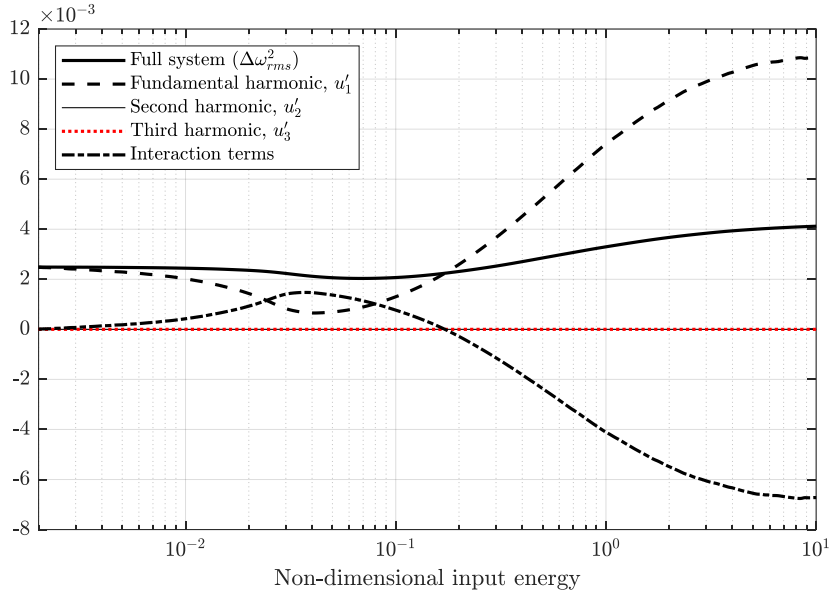

Figure 13. Contribution of each term in the bandwidth-squared formula, (17). Comparison between the contribution of the fundamental (black dashed curve), second (thin black solid curve), third (red dotted curve) and the inter-harmonic interactions term (black dash-dotted curve) with $\Delta\omega_{rms}^2$ of the original system, (13), with $\phi_0 = 8°$.

That the inter-harmonic interaction term is negative in the hardening regime indicates that the higher harmonics feed off energy from the fundamental harmonic (like the previous, purely hardening case); hence, there occurs nonlinear targeted energy transfer from low-to-high frequencies. In addition, the fact that in this regime the fundamental harmonic bandwidth contribution exceeds the overall bandwidth $\Delta\omega_{rms}^2$ leads to the same conclusion, i.e., that the energy loss rate of the fundamental harmonic is slower compared to the full system. Indeed, not all the energy loss of the fundamental harmonic is attributed to viscous dissipation, since part of it is transferred to the higher harmonics, which increases the overall energy decay rate of the oscillator response (since energy is dissipated faster by the higher harmonics). A different picture, however, is observed at the intermediate-to-low energy softening regimes, where the inter-harmonic interaction term reverses sign and is now positive; moreover, the contribution to the bandwidth of the fundamental harmonic drops below the overall bandwidth $\Delta\omega_{rms}^2$. This indicates that in the softening regime the higher harmonics transfer energy to the fundamental harmonic, which decreases the overall rate of energy dissipation (since the lower frequency fundamental harmonic dissipates energy more slowly compared to the higher harmonics). This results in a decrease in the bandwidth in that regime. Lastly, as the energy to the system asymptotically reaches zero, system (13) becomes linearized, the nonlinear higher harmonics become negligible, and the transient dynamics is dominated by the fundamental harmonic. This is evident in the wavelet and inverse wavelet transforms of Fig. 13 and the asymptotic low energy decay of the inter-harmonic interaction contribution to the bandwidth in Fig. 14.

In addition to the systems studied in this and the previous sections, the bandwidth of a SDOF passive system with inelastic single-sided vibro-impact nonlinearity and weak Coulomb friction was computationally and experimentally studied in a recent study [39]. It was shown that the combined effects of the single-sided nonlinearity (which is a hardening type pf nonlinearity), impact elasticity and Coulomb friction, significantly *increase* the bandwidth of the system. Moreover, by slightly adjusting the configuration of the system, i.e., by carefully placing pairs of magnets, to create an effective softening stiffness nonlinearity, the authors showed that the dissipative effects of inelastic impacts and friction effects can be overcome and the bandwidth of the system can be *decreased*. These results were used to computationally and experimentally prove that unlike SDOF *linear* time-invariant (LTI) systems which have time-bandwidth products of unity (known as the "time-bandwidth limit" [40]), this passive nonlinear system can possess time-bandwidth products of greater and less than unity, depending on the configuration of system, i.e., employing softening and hardening nonlinearity, respectively.

## 5. Concluding remarks

The concept of bandwidth as employed across the sciences and engineering is used to characterize the dissipative capacity of a system in the frequency and time domains and is relevant over a broad range of fields and applications. Because it is used in various contexts and communities, different definitions have evolved to accommodate diverse theories and analyses. Accordingly, the focus in this article is a general definition for the bandwidth for a broad class of linear and nonlinear, single and multiple degree-of-freedom, time-invariant and -variant oscillators. Traditionally, the bandwidth of a SDOF oscillator is determined by the half-power (-3 dB) method under certain restrictions which include linearity, weak dissipation, and stationary output; moreover, it applies to SDOF oscillators and to MDOF oscillators with modes well separated in frequency. In this work all these restrictions are relaxed by proposing a definition based on the root mean square (RMS) bandwidth and the envelope of the decaying energy of the oscillator. This relates to the original purpose of bandwidth, which is to quantify the overall dissipative capacity of the system or, equivalently, to describe how localized the energy of the system is in the time and frequency domains. Since the RMS bandwidth of the energy signal is linearly proportional to the inverse of the variance of the energy signal in the time domain (according to the Fourier uncertainty principle), the new bandwidth definition provides an accurate measure of the dissipation rate (capacity) of the free decay of the oscillator and is an inherent property. Moreover, employing the envelope of the energy in the bandwidth computation eliminates complexities associated with time-varying frequencies, i.e., signal non-stationarity, and readily accommodates multi-mode oscillating systems.

The applications discussed showed that the new bandwidth definition is not only in agreement with the traditional half-power bandwidth definition in the linearized regime of the transient dynamics of a nonlinear oscillator, but also accurately captures the qualitative features, e.g., softening, hardening, or bistability, of the nonlinear dynamics at high-to-intermediate energy levels. Specifically, it was demonstrated how such strongly nonlinear features affect the dissipative capacity of the oscillator, yielding a decrease in the bandwidth (compared to the linearized half-power measure) in softening dynamical regimes, or bandwidth increase in hardening regimes. In addition, it was shown that bistability can, under certain conditions, cause a drastic increase (in the studied example, as much as 600%) or decrease (by as much as 12 %) of the oscillator bandwidth compared to the linearized bandwidth limit.

Furthermore, by expressing the bandwidth in terms of the contributions of the fundamental and higher harmonics that are generated by the nonlinearity in the transient response, and employing inverse continuous wavelet transform harmonic decompositions, it was possible to quantify the contribution of each individual harmonic to the bandwidth, and, perhaps, even more interesting, the bandwidth contributions of inter-harmonic interactions, i.e., of targeted energy transfers between different harmonics. A typical trend was that, individually, only the fundamental harmonic contributes in a significant way to the bandwidth. The individual contributions to the bandwidth of the higher harmonics are negligible, since these harmonics get their energy by nonlinear scattering of energy from the fundamental harmonic in the first place, so unlike the dominant harmonic they do not has their own individual energy, but rather gain of lose energy through interactions with the dominant harmonic and other harmonics. However, once energy gets nonlinear scattered to the higher harmonics, targeted energy transfers from the fundamental harmonic to the higher harmonics in regimes of hardening nonlinearity increased the bandwidth, whereas, the reverse, i.e., directed energy transfers from higher harmonics to the fundamental harmonic in regimes of softening nonlinearity decreased the bandwidth. This occurs since transferring energy from the (lower) fundamental harmonic to the higher harmonics enhances the overall dissipative capacity of the oscillator (yielding an increase in the bandwidth), whereas transferring energy from the higher harmonics to the fundamental one decreases the overall dissipation rate (resulting in a bandwidth decrease).

These findings demonstrate that the new bandwidth definition represents a generalization of the traditional linear half-power bandwidth and may find broad application in the sciences and engineering.

36. Mojahed A, Liu Y, Bergman LA *et al.* (2021) Modal energy exchanges in an impulsively loaded beam with a geometrically nonlinear boundary condition: computation and experiment. *Nonlinear Dynamics* **103**, 3443-3463.

37. Mojahed A, Bergman LA, Vakakis AF (2020) Tunable-with-energy intense modal interactions induced by geometric nonlinearity. *Proceedings of the Institution of Mechanical Engineers, Part C: Journal of Mechanical Engineering Science*, 0954406220908621.

38. Mojahed A, Bergman LA, Vakakis AF (2021) New inverse wavelet transform method with broad application in dynamics. *Mechanical Systems and Signal Processing* **156**, 107691.

39. Mojahed A, Tsakmakidis KL, Bergman LA, Vakakis AF (2021) Exceeding the Classical Time-bandwidth Product in Nonlinear Time-invariant Systems. *Nonlinear Dynamics* (In review).

40. Born M. (1961) Bemerkungen zur statistischen Deutung der Quantenmechanik. InWerner Heisenberg und die Physik unserer Zeit. *Vieweg+ Teubner Verlag, Wiesbaden*, 103-118.